\documentclass[showpacs,bibnotes,pra,twocolumn,superscriptaddress]{revtex4}%
\usepackage{epsfig}
\usepackage{graphicx}
\usepackage{bm}
\usepackage{amsmath}
\usepackage{graphicx}
\usepackage{amsfonts}
\usepackage{amssymb}%
\begin{document}
\title{Quantum optomechanics beyond linearization}
\author{Bing He}
\affiliation{University of California, Merced, 5200 North Lake Road, Merced, CA 95343, USA}

\pacs{42.50.Pq, 37.30.+i, 42.50.Wk}

\begin{abstract}
The quantum dynamics of optomechanical systems was mostly studied for their fluctuations around classical steady states. We present a theoretical approach to determining the system observables of optomechanical systems as genuine quantum objects, for example, a coupled quantum mechanical oscillator to a cavity single photon. In this approach we study the dynamics of such systems in strong coupling regime. We find that, under strong optomechanical coupling, steady quantum states of optomechanical systems driven by continuous-wave single photons exhibit periodic oscillation and cavity noise considerably affects system observables.
\end{abstract}
\maketitle

\section{Introduction}
Optomechanical systems (OMS) provide a platform to study macroscopic quantum phenomena \cite{R1,R2}.
Recent experiments have achieved the motional ground states of micro-nano oscillators \cite{ex00, ex01, ex02}, creating the possibility toward manipulating macroscopic quantum objects. On the other hand, experimental systems are approaching the single-photon strong coupling regime \cite{ex1,ex2,ex3}, where the radiation pressure of a single photon could displace mechanical oscillator by more than its zero-point uncertainty. Strong optomechanical coupling is useful to making the macroscopic superposition states proposed for testing quantum theory \cite{p1,p2,p3}. 

The theoretical understanding on this novel regime of single photon and strong coupling optomechanics is under way too. The initial studies by Rabl \cite{s1}, Nunnenkamp and co-workers \cite{s2}, apply the linearized Langevin equation and/or the master equation about quantum fluctuations around classical steady states to find system observables. Another existing theoretical approach adopts the Schr\"{o}dinger equation under the pure quantum state assumption which is valid in single-photon subspace and for negligible loss of mechanical oscillator \cite{s3,s4}. Other study involving strong optomechanical coupling can be found in \cite{l-k-f}. Despite the progress, more appropriate treatment should be developed for quantum OMS that can not be simply described as classical steady state plus quantum fluctuation to have linearized equations of motion.

To have a clearer picture of the problem, we refer to the approach in the previous studies. 
This is the linearization of the Langevin equation from the Hamiltonian ($\hbar\equiv 1$)
\begin{eqnarray}
\hat{H}&=&\omega_c\hat{a}^{\dagger}\hat{a}+\omega_m\hat{b}^{\dagger}\hat{b}-g(\hat{b}+\hat{b}^{\dagger})\hat{a}^{\dagger}\hat{a}\nonumber\\
&+&iE(\hat{a}^{\dagger}e^{-i\omega_0t}-\hat{a}e^{i\omega_0t})
\label{total-hamiltonian}
\end{eqnarray}
of a generic OMS, where $g$ is the optomechanical coupling intensity, $E$ the driver intensity, and $\omega_c$, $\omega_m$ the cavity resonance frequency, mechanical frequency, respectively. Here we consider 
a continuous-wave driver with the central frequency $\omega_0$.
In a frame rotating with the driver frequency $\omega_0$ and diagonalizing the system Hamiltonian (the terms except
for the driving term in (\ref{total-hamiltonian})), the exact Langevin equation for the cavity mode reads
\vspace{-0.1cm}
\begin{equation}
\dot{\hat{a}}=-\frac{\kappa}{2}\hat{a}+i\Delta \hat{a}+i\frac{2g^2}{\omega_m}\hat{a}^{\dagger}\hat{a}\hat{a}+e^{-i\hat{Q}_m}(E-\sqrt{\kappa}\hat{\xi}_c),
\label{langevin}
\vspace{-0.1cm}
\end{equation}
where $\Delta=g^2/\omega_m-\Delta_0$ with $\Delta_0=\omega_c-\omega_0$ being the detuning of the driver frequency  from the cavity frequency, $\hat{Q}_m=ig/\omega_m(\hat{b}^{\dagger}-\hat{b})$, and $\kappa$, $\hat{\xi}_c$ the cavity damping rate and cavity noise operator, respectively. The effect of the nonlinear term, the third term on the right-hand side of (\ref{langevin}), is canceled with the argument of weak driver in \cite{s1,s2}. In fact it is to perform a displacement $\hat{a}\rightarrow \alpha_s+\hat{a}$, where the classical steady state value $\alpha_s$ of the cavity field is proportional to the driver intensity $E$, so that Eq. (\ref{langevin}) will be reduced to a tractable linear equation neglecting the higher order terms of $E$. The results based on such linearization are consistent with the numerical calculation with the master equation about the fluctuations around classical cavity steady states \cite{s2}, showing its equivalence to the practice of decomposing the system operators into classical steady state values plus quantum fluctuations (see, e.g. \cite{vitali}) adopted in most previous studies on OMS. However, in the situations of a cavity single photon coupled to a mechanical oscillator in quantum regime, there is no way to define a classical steady state for the single quanta $\hat{a}$ since it is a quantum fluctuation itself. Then, for the single photon optomechanics, the effect of the nonlinear term in (\ref{langevin}) should not be simply neglected to the dynamics of OMS, and any linearization for this type of problems is not appropriate. 

In this work we present an approach that models OMS as quantized interacting oscillators (one cavity and one mechanical mode) which damp energy into their environment and are subject to quantum noise effects at the same 
time. This full quantum approach allows one to determine the system dynamics without the reliance on the classical configurations of OMS, because it is based on the evolution of quantum operators. Though the approach applies to the general coupling of OMS, we will primarily study the physics of quantum OMS in strong coupling regime by illustrating the evolution of the relevant system observables and clarifying the physical factors to determine their values. 
Moreover, the effects of quantum noise can be well captured in the approach. 

\section{System evolution and observables}
We start with the quantum state evolution of a generic OMS governed by the master equation ($\hbar\equiv 1$)
\begin{eqnarray}
\dot{\rho}&=&
-i[\hat{H}_S(t),\rho]+\kappa{\cal D}[\hat{a}]\rho+\gamma_m(n_{th}+1){\cal D}[\hat{b}]\rho\nonumber\\
&+&\gamma_m n_{th}{\cal D}[\hat{b}^{\dagger}]\rho
\equiv  {\cal L}(t)\rho
\label{master}
\end{eqnarray}
with the system Hamiltonian \cite{H}
\begin{eqnarray}
\hat{H}_S(t)
&=& -\sqrt{2}g\hat{a}^{\dagger}\hat{a}\{\hat{x}_m\cos(\omega_mt)+\hat{p}_m\sin(\omega_mt)\}\nonumber\\
&+&iE(e^{i\Delta_0 t}\hat{a}^{\dagger}-e^{-i\Delta_0 t}\hat{a})
\label{om}
\end{eqnarray}
in the interaction picture (the purpose for adopting the interaction picture will be seen below), where $n_{th}=(e^{\omega_m/k_BT}-1)^{-1}$ is the thermal phonon number at the temperature $T$, and $\kappa$, $\gamma_m$ the cavity and mechanical damping rate, respectively. Here we define $\hat{x}_m=(\hat{b}+\hat{b}^{\dagger})/\sqrt{2}$ and  $\hat{p}_m=-i(\hat{b}-\hat{b}^{\dagger})/\sqrt{2}$ as the dimensionless displacement and momentum operator of the mechanical oscillator with the frequency $\omega_m$. ${\cal D}[\hat{c}]\rho=\hat{c}\rho\hat{c}^{\dagger}-(\rho\hat{c}^{\dagger}\hat{c}+\hat{c}^{\dagger}\hat{c}\rho)/2$ is the dissipator in Lindblad form. The initial state of OMS is prepared as $\rho(0)=|0\rangle_c \langle 0|\otimes \rho_{th}$, the tensor product of cavity vacuum and mechanical thermal state. As in solving the Langevin equation in (\ref{langevin}), the nonlinear coupling term in (\ref{om}) makes it difficult to find the analytical solution to the master equation in (\ref{master}).

Here we provide a method to calculate the expectation value $\langle \hat{O}_S\rangle=Tr_S \{\hat{O}_S \rho(t)\}$  of a system operator $\hat{O}_S $ without the solution $\rho(t)$ to the master equation in (\ref{master}).
To do this, we look at the combinations of OMS and their reservoirs. The damping of cavity field and mechanical oscillator can be described in terms of a linear coupling of the system operators $\hat{c}_i=\hat{a}, 
\hat{b}$ with the quantum noise operators $\hat{\xi}_c$, $\hat{\xi}_m$ of the cavity and mechanical reservoir, respectively. In this picture the overall evolution for the combination of an OMS and its associated reservoirs is determined by the following unitary evolution operator of stochastic Hamiltonian \cite{b1}
\begin{eqnarray}
\hat{U}(t,0)&=& \mbox{T}\exp\big\{-i\int_0^t\hat{H}_S(\tau) d\tau \nonumber\\
&+&\sum_i  \sqrt{\gamma_i}\int_0^t(d\hat{B}_i^{\dagger}(\tau)\hat{c}_i
- d\hat{B}_i(\tau)\hat{c}_i^{\dagger})\big\},
\label{U}
\end{eqnarray}
where $\mbox{T}$ stands for a time-ordered operation, $\hat{B}_i(t)=\int_0^t \hat{\xi}_i (\tau)d\tau$, $\gamma_i=\kappa,\gamma_m$, and $\hat{H}_S(t)$ takes the form in the interaction picture. The stochastic operators $d\hat{B}_i$ and their conjugates satisfy the Ito's rules corresponding to the correlators $\langle \hat{\xi}_c(t)\hat{\xi}^{\dagger}_c(t')\rangle_R=\delta(t-t')$, $\langle \hat{\xi}^{\dagger}_c(t)\hat{\xi}_c(t')\rangle_R=0$ of the cavity vacuum noise, and $\langle \hat{\xi}_m(t)\hat{\xi}^{\dagger}_m(t')\rangle_R=(n_{th}+1)\delta(t-t')$, $\langle \hat{\xi}^{\dagger}_m(t)\hat{\xi}_m(t')\rangle_R=n_{th}\delta(t-t')$ of the mechanical thermal bath; 
see Ref. \cite{b1} for the details. Tracing out the reservoir degrees of freedom in the increment by infinitesimal transformation, $\hat{U}(t+dt,t)\hat{\rho}(t)\hat{U}^{\dagger}(t+dt,t)-\hat{\rho}(t)$, restores the master equation in (\ref{master}). The density matrix $\hat{\rho}(t)$ for the combination of OMS and reservoir is assumed to factorize
at $t=0$, i.e. $\hat{\rho}(0)=\rho(0)\otimes \rho_R$ with $\rho_R$ being the tensor product of cavity reservoir vacuum state and mechanical reservoir thermal state. 
The expectation value of a system operator can be rewritten with $\hat{U}(t,0)$ as 
\begin{eqnarray}
&&Tr_S \{\hat{O}_S \rho(t)\}\nonumber\\
&=&Tr_S \{\hat{O}_S \hat{U}_0(t,0)Tr_R \big(\hat{U}(t,0)\rho(0)\otimes\rho_R\hat{U}^{\dagger}(t,0)\big)\hat{U}^{\dagger}_0(t,0)\}\nonumber\\
&=&Tr_{S,R}\{\hat{U}^{\dagger}(t,0)\hat{U}^{\dagger}_0(t,0)\hat{O}_S\hat{U}_0(t,0)\hat{U}(t,0)\rho(0)\otimes \rho_R\}, 
\end{eqnarray}
where $\hat{U}_0(t,0)=\exp\{-i(\omega_c \hat{a}^{\dagger}\hat{a}+\omega_m \hat{b}^{\dagger}\hat{b})t\}$ 
converts the state in the interaction picture to that of the Schr\"{o}dinger picture. 

For any operator $\hat{O}_S=\hat{f}(\hat{c}_i,\hat{c}^{\dagger}_i)$, the transformation by $\hat{U}_0(t,0)$ only adds the phase $e^{-i\omega_it}$ ($e^{i\omega_it}$), where $i=c$ or $m$, to $\hat{c}_i$ ($\hat{c}^{\dagger}_i$). Its expectation value as a system observable is therefore the average of the transformed functional operator $\hat{U}^{\dagger}(t,0)\hat{f}(\hat{c}_i,\hat{c}^{\dagger}_i)\hat{U}(t,0)=
\hat{f}\big(\hat{U}^{\dagger}(t,0)\hat{c}_i\hat{U}(t,0),\hat{U}^{\dagger}(t,0)\hat{c}^{\dagger}_i\hat{U}(t,0)\big)$ over the total initial state of system plus reservoir, with the above-mentioned phase absorbed in 
$\hat{c}_i$ ($\hat{c}^{\dagger}_i$). The reduction of determining system observables to finding the transformed basic system operators $\hat{U}^{\dagger}(t,0)\hat{c}_i\hat{U}(t,0)$ is the main advantage of our combined unitary evolution approach. The more complicated averages such as $\langle \hat{O}_1(t+\tau)\hat{O}_2(\tau)\rangle$, which is generally calculated with the quantum regression formula \cite{b2}, can be calculated in a similar way.

\section{Decomposition of system-reservoir evolution}
We will find the expectation value of a system operator $\hat{O}_S=\hat{f}(\hat{c}_i,\hat{c}^{\dagger}_i)$ by averaging its transformation $\hat{f}\big(\hat{U}^{\dagger}(t,0)\hat{c}_i\hat{U}(t,0),\hat{U}^{\dagger}(t,0)\hat{c}^{\dagger}_i\hat{U}(t,0)\big)$ over the initial system state plus reservoir state. The total unitary evolution operator $\hat{U}(t,0)$, however, involves the driving on cavity, the optomechanical coupling, as well as the coupling between system and reservoirs, which are not mutually commutative quantum processes. To simplify the calculation of system operator transformations under such overall unitary evolution, we will need the following decompositions 
\begin{eqnarray}
&&\mbox{T}e^{-i\int_0^t d\tau (\hat{H}_1(\tau)  +\hat{H}_2 (\tau))}\nonumber\\
&=&\mbox{T}e^{-i\int_0^t d\tau \hat{V}_2(t,\tau )\hat{H}_1(\tau)\hat{V}^{\dagger}_2(t,\tau) }~\mbox{T}e^{-i\int_0^t d\tau \hat{H}_2(\tau)},
\label{1}
\end{eqnarray}
where $\hat{V}_2(t,\tau )=\mbox{T}\exp\{-i\int_\tau^t d\tau' \hat{H}_2 (\tau')\}$, and
\begin{eqnarray}
&&\mbox{T}e^{-i\int_0^t d\tau (\hat{H}_1(\tau) +\hat{H}_2 (\tau))}\nonumber\\
&=& \mbox{T}e^{-i\int_0^t d\tau \hat{H}_1(\tau)  }~\mbox{T}e^{-i\int_0^t d\tau \hat{V}^{\dagger}_1(\tau,0)\hat{H}_2(\tau)\hat{V}_1(\tau,0)},
\label{2}
\end{eqnarray}
where $\hat{V}_1(\tau,0)=\mbox{T}\exp\{-i\int_0^\tau d\tau' \hat{H}_1(\tau')\}$, of a time-ordered exponential. The proof of the two decompositions is given in Appendix A.

First, applying (\ref{1}) to (\ref{U}), we separate out a unitary evolution operator $\hat{V}_D(t,0)$ of the system-reservoir coupling so that the overall unitary evolution operator can be decomposed into the form $\hat{U}(t,0)=
\hat{V}_{S}(t,0)\hat{V}_D(t,0)$, where $\hat{V}_D(t,0)=\mbox{T}\exp \{\sum_i(\int_0^t \sqrt{\gamma_i}d\hat{B}_i^{\dagger}(\tau)\hat{c}_i- \int_0^t\sqrt{\gamma_i}d\hat{B}_i(\tau)\hat{c}_i^{\dagger})\}$ describes the coupling between the system and reservoirs. 
The unitary operation $\hat{V}_D(t,\tau )$ inside $\hat{V}_{S}(t,0)=\mbox{T}\exp\{-i\int_{0}^t d\tau\hat{V}_D(t,\tau )\hat{H}_S(\tau)\hat{V}^{\dagger}_D(t,\tau)\}$ transforms the system operators 
$\hat{c}_i=\hat{a}$, $\hat{b}$ in $\hat{H}_S(\tau)$ to 
\begin{eqnarray}
\hat{V}_D(t,\tau )\hat{c}_i\hat{V}^{\dagger}_D(t,\tau)&=&e^{-\gamma_i(t-\tau)/2}\hat{c}_i+\hat{n}_i(t,\tau)\equiv \hat{c}_i(t,\tau),\nonumber\\
\label{1st}
\end{eqnarray}
with 
$\hat{n}_i(t,\tau)=\sqrt{\gamma_i}\int_{\tau}^t e^{-\gamma_i(t'-\tau)/2}\hat{\xi}_i(t')dt'$ being
the induced quantum colored noise operators satisfying the commutation relation
\begin{eqnarray}
\Gamma_i(\tau,\tau')&=&[\hat{n}_i(t,\tau),\hat{n}_i^{\dagger}(t,\tau')]\nonumber\\
&=&e^{-\gamma_i|\tau-\tau'|/2}-e^{-\gamma_i(t-\tau)/2}e^{-\gamma_i(t-\tau')/2};\nonumber\\
\label{correlation}
\end{eqnarray}
see Appendix B for the proof. From now on, one will rewrite the expectation value of a system operator as
\begin{eqnarray}
Tr_S \big(\hat{O}_S\rho(t)\big) &=& Tr_{S,R}\big \{\hat{V}^{\dagger}_S(t,0)\hat{U}_0^{\dagger}(t,0)\hat{O}_S \hat{U}_0(t,0)\hat{V}_S(t,0)\nonumber\\
&\times & \hat{V}_D(t,0)\rho(0)\otimes \rho_R \hat{V}^{\dagger}_D(t,0)\big\}.
\label{1-sim}
\end{eqnarray}
This expectation value will be therefore determined by the 
transformations $\hat{V}_S^{\dagger}(t,0)\hat{c}_i\hat{V}_S(t,0)$, together with $\hat{V}_D^{\dagger}(t,0)\hat{c}_i\hat{V}_D(t,0)$ that are similar to Eq. (\ref{1st}) \cite{ep}.

Next, using Eq. (\ref{2}), we will separate the pure driving process and the optomechanical process in the unitary evolution operator $\hat{V}_S(t,0)=\mbox{T}\exp\{-i\int_0^t d\tau \hat{H}'_S(t,\tau)\}$, where $\hat{H}'_S(t,\tau)=iE \{\hat{a}^{\dagger}(t,\tau)e^{i\Delta_0 \tau}-\hat{a}(t,\tau)e^{-i\Delta_0 \tau}\}
- g\hat{K}_m(t,\tau)\hat{a}^{\dagger}(t,\tau)\hat{a}(t,\tau)$, to have the product $\hat{V}_S(t,0)=\hat{V}_E(t,0)\hat{V}_{OM}(t,0)$. The pure driving operator, $\hat{V}_E(t,0)= \mbox{T}\exp\big \{E\int_0^t d\tau  (\hat{a}^{\dagger}e^{i\Delta_0 \tau}e^{-\frac{\kappa}{2}(t-\tau)}-h.c.)\big\}\times \mbox{T}\exp\big \{E\int_0^t d\tau  (e^{i\Delta_0 \tau}\hat{n}^{\dagger}_c(t,\tau)-h.c.)\big\}$, consists of the actions by the external driver alone and the noise $\hat{n}_c(t,\tau)$ from the vacuum reservoir. The form of the optomechanical coupling evolution operator $\hat{V}_{OM}(t,0)=\mbox{T}\exp \big\{ig\int_0^t d\tau \hat{K}_m(t,\tau)\hat{C}^{\dagger}(t,\tau)\hat{C}(t,\tau)\big\}$ is obtained by (\ref{1}) and (\ref{2}), from which we have
$\hat{K}_m(t,\tau)=\cos(\omega_m \tau)\hat{X}_m(t,\tau)+ \sin(\omega_m \tau)\hat{P}_m(t,\tau)$ with
\begin{eqnarray}
\hat{X}_m(t,\tau)
&=& \sqrt{2}e^{-\frac{\gamma_m}{2}(t-\tau)}\hat{x}_m+(\hat{n}_m(t,\tau)+h.c.)\nonumber\\
\hat{P}_m(t,\tau)
&=& \sqrt{2}e^{-\frac{\gamma_m}{2}(t-\tau)}\hat{p}_m-(i\hat{n}_m(t,\tau)-h.c.)~~~
\label{x-p}
\end{eqnarray}
being the transformed displacement and momentum operator of the mechanical oscillator, as well as the transformed cavity operator
\begin{eqnarray}
\hat{C}(t,\tau)&=&\hat{V}_E^{\dagger}(\tau,0)\hat{V}_{D}(t,\tau)\hat{a}\hat{V}^{\dagger}_{D}(t,\tau)\hat{V}_E(\tau,0)\nonumber\\
&=&e^{-\frac{\kappa}{2}(t-\tau)}\big(\hat{a}+D_1(\tau)\big)+\hat{n}_c(t,\tau)+D_2(\tau)\nonumber\\
&\equiv &e^{-\frac{\kappa}{2}(t-\tau)}\hat{a}+\hat{G}(t,\tau).
\label{t-ca}
\end{eqnarray} 
In (\ref{t-ca}), the functions due to the displacement by the external driver and cavity noise operator in $\hat{V}_E(\tau,0)$ are respectively found as  
\begin{eqnarray}
D_1(\tau)&=&
(e^{-\kappa(t-\tau)/2+i\Delta_0\tau}
- e^{-\kappa t/2}) \frac{E}{\frac{\kappa}{2}+i\Delta_0},\nonumber\\
D_2(\tau)&=& E\int_0^\tau d t'~e^{i\Delta_0 t'}\Gamma_c(t',\tau).
\end{eqnarray}
\vspace{0.2cm}

The unitary evolution operator for the combination of OMS and reservoirs has now been decomposed as $\hat{U}(t,0)=\hat{V}_E(t,0)\hat{V}_{OM}(t,0)\hat{V}_D(t,0)$. Such decomposition of a quantum physical process 
into three dependent but simplified ones makes it possible to find the transformations $\hat{U}^{\dagger}(t,0)\hat{c}_i\hat{U}(t,0)$ of the basic OMS operators. As it will be shown below, the contributions to a system observable from the different factors, e.g., the external driving and the optomechanical coupling, can be seen by such decomposition as well. It is therefore convenient for the approach to study the system dynamics in various different regimes, as one adjusts the system parameters to the decomposed processes.

\begin{figure}[b!]
\vspace{-0cm}
\centering
\epsfig{file=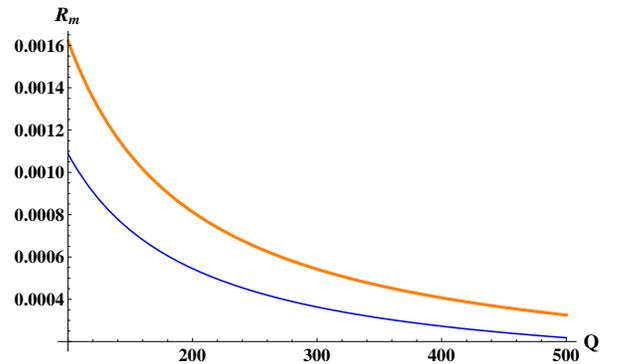,width=0.9\linewidth,clip=} 
{\vspace{-0cm}\caption{\label{Fig:c} (color online) Proportion of mechanical noise corrected cavity photon number $R_m=\Delta n_P/n_P$ v.s. quality factor $Q=\omega_m/\gamma_m$. 
The thick solid (orange) line is for the resolved-sideband parameter $\omega_m/\kappa=1$, and the thin solid (blue) line for $\omega_m/\kappa=2$. The parameters are $g/\kappa=0.5$, $\Delta_0/\omega_m=0$, and $T=0$. The plots show the mechanical noise correction at $\kappa t=10$.}}
\vspace{-0.5cm}
\end{figure}

\section{Dynamical and noise effects}
Now we go back to study the physics of quantum OMS with the above decomposition technique.
The infinite product expression for the unitary evolution operator $\hat{V}_{OM}(t,0)$ in $\hat{V}_S(t,0)=\hat{V}_E(t,0)\hat{V}_{OM}(t,0)$ enables one to obtain an analytical form of the transformation $\hat{V}^{\dagger}_{S}(t,0)\hat{a}\hat{V}_{S}(t,0)$ (see Appendix C for the details). For a weak drive of $E/\kappa\ll 1$, its average over the initial cavity vacuum state reads 
\begin{widetext}
\vspace{-0.8cm}
\begin{eqnarray}
\langle \hat{V}^{\dagger}_{S}(t,0) \hat{a}\hat{V}_S(t,0)\rangle_c 
&=& D_1(t)
+ ig \int_0^t d\tau ~ \exp\{-ig^2\Theta(\tau)\}\exp\{ig\int_0^\tau du~ e^{-\kappa (t-u)}\sin(\omega_m u)\hat{P}_m(t,u)\}\nonumber\\
&\times &\exp\{ig\int_0^\tau du~ e^{-\kappa (t-u)}\cos(\omega_m u)\hat{X}_m(t,u)\}\hat{K}_m(t,\tau)
e^{-\kappa(t-\tau)/2}\hat{G}_C(t,\tau),
\label{cavity}
\end{eqnarray}
where
\begin{eqnarray}
\hat{G}_C(t,\tau)&=&\hat{G}(t,\tau)+ig\int_{\tau}^t d\tau_1 \Gamma_c(\tau_1,\tau)\hat{K}_m(t,\tau_1)\hat{G}(t,\tau_1)+(ig)^2 \int_{\tau}^t d\tau_1 \Gamma_c(\tau_1,\tau)\hat{K}_m(t,\tau_1)\int_{\tau_1}^t d\tau_2 \Gamma_c(\tau_2,\tau_1)\hat{K}_m(t,\tau_2)\hat{G}(t,\tau_2)\nonumber\\
&+& (ig)^3 \int_{\tau}^t d\tau_1 \Gamma_c(\tau_1,\tau)\hat{K}_m(t,\tau_1)\int_{\tau_1}^t d\tau_2 \Gamma_c(\tau_2,\tau_1)\hat{K}_m(t,\tau_2)
\int_{\tau_2}^t d\tau_3 \Gamma_c(\tau_3,\tau_2)\hat{K}_m(t,\tau_3)\hat{G}(t,\tau_3)+\cdots
\label{drive}
\end{eqnarray}
\vspace{-0.3cm}
\end{widetext}
includes the correction to the drive operator $\hat{G}(t,\tau)$ defined in (\ref{t-ca}) by the induced cavity colored noise $\hat{n}_c$ with the correlation function $\Gamma_c(\tau,\tau')$. The extra phase
$\Theta(\tau)=2\int_0^\tau du e^{-\kappa (t-u)}\sin(\omega_m u)\int_0^{u} dv e^{-\kappa(t-v)-\gamma_m (u-v)/2}\cos(\omega_m v)$ is due to the non-commutativity between $\hat{X}_m(t,\tau)$ and $\hat{P}_m(t,\tau)$.
The term $D_1(t)=E(e^{i\Delta_0 t}
- e^{-\kappa t/2})/(\frac{\kappa}{2}+i\Delta_0)$ simply arises from the pure driving process $\hat{V}_E(t,0)$.

\subsection{Weak coupling limit}

In the weak coupling limit $g\ll \kappa$, the average $\langle \hat{V}^{\dagger}_{S}(t,0) \hat{a}\hat{V}_S(t,0)\rangle_c $ could be approximated by the term
\begin{eqnarray}
\hat{l}(t)&=&\frac{gE}{\frac{\kappa}{2}+i\Delta_0}(i\frac{ \hat{x}_m}{\sqrt{2}}+\frac{\hat{p}_m}{\sqrt{2}})\big (\frac{e^{i(\Delta_0+\omega_m)t}}{i(\Delta_0+\omega_m)+(\frac{\kappa}{2}+\frac{\gamma_m}{2})} \nonumber\\
& +&\frac{e^{i(\Delta_0-\omega_m)t}}{i(\Delta_0-\omega_m)+(\frac{\kappa}{2}+\frac{\gamma_m}{2})} \big) 
\label{1-st}
\end{eqnarray}
in the lowest $g$ order of Eq. (\ref{cavity}),
in addition to the pure driving and noise correction terms. Showing the anti-Stokes and Stokes sidebands, this approximation well explains the physics of OMS in linearized and weak-coupling regime. 

An important feature in our approach is that a system observable should be 
found as the average of a transformed system operator over the initial state of both OMS and reservoir. The average over reservoir state manifests the mechanical and cavity noise corrections to the system observable.
For instance, the colored mechanical noise operator $\hat{n}_m$ in $\hat{X}_m(t,\tau)$ and $\hat{P}_m(t,\tau)$ of (\ref{cavity}) corrects the mean cavity photon number $n_P=\langle\hat{a}^{\dagger}\hat{a}\rangle$. In the lowest order, which well describes the weak couping regime, it changes the photon number by 
\begin{eqnarray}
\Delta n_P&=&\frac{g^2E^2}{\frac{\kappa^2}{4}+\Delta^2_0}\int_0^t d\tau_1\int_0^t d\tau_2 e^{-\kappa(t-\tau_1)/2}e^{-\kappa(t-\tau_2)/2}\nonumber\\
&\times & \cos\omega_m(\tau_1-\tau_2)e^{i\Delta_0\tau_1-i\Delta_0\tau_2}\Gamma_m(\tau_1,\tau_2)
\end{eqnarray} 
at $T=0$ (extra terms containing $n_{th}$ will be added for a system at the temperature $T>0$). The corrections in the higher orders can be calculated with Wick's theorem to sum up the products of $\langle \hat{n}^{\dagger}_m(t,\tau)\hat{n}_m(t,\tau')\rangle_R$ and $\langle \hat{n}_m(t,\tau)\hat{n}^{\dagger}_m(t,\tau')\rangle_R$. The contribution of these correlators to photon number $n_P$ is rather small under the condition $\gamma_m\ll \kappa$. Fig. 1 illustrates the proportion of such correction, $R_m=\Delta n_P/n_P$ where $n_P=\langle (\hat{l}^{\dagger}+D_1^{\ast})(\hat{l}+D_1)\rangle+\Delta n_P$, in the total mean cavity photon number of the weak coupling regime, showing that $\Delta n_P$ will become more negligible with the increasing quality factor of mechanical oscillator. 

\subsection{Transition from weak coupling to strong coupling regime}

\begin{figure}[t!]
\vspace{-0cm}
\centering
\epsfig{file=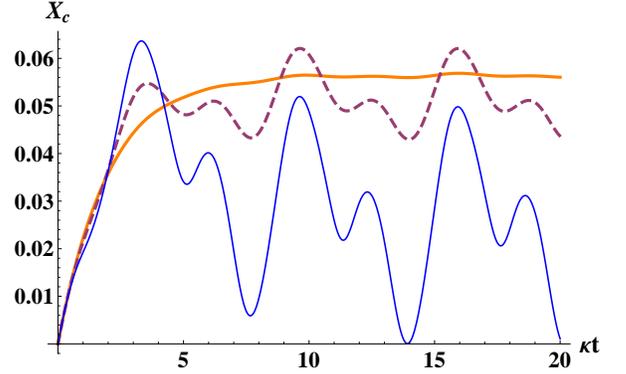,width=0.9\linewidth,clip=} 
{\vspace{-0cm}\caption{\label{Fig:q} (color online) Real-time evolution of cavity quadrature $X_c$ with the dimensionless time $\kappa t$. The thick solid (orange) line shows the process for the coupling $g/\kappa=0.1$, the dashed line for $g/\kappa=0.5$, and the thin solid (blue) line for $g/\kappa=1$. Here we choose the system parameters to be $E/\kappa=0.01$, $\omega_m/\kappa=1$, $\omega_m/\gamma_m=100$, $\Delta_0/\omega_m=0$, and $T=0$. }}
\vspace{-0.1cm}
\end{figure}

Before we discuss the cavity noise effect, we take a look at the effect of optomechanical coupling on cavity observables as an OMS undergoes the transition from weak coupling to strong coupling regime. Here we take the cavity quadrature $X_c=\langle \hat{a}+\hat{a}^{\dagger}\rangle/\sqrt{2}$ for illustration. From Eq. (\ref{cavity}) the further averages are taken over the initial mechanical oscillator and reservoir state to obtain the numerical values of $X_c$.  Fig. 2 shows the real-time evolution of $X_c$ for three different optomechanical coupling intensities. For the weakest coupling, the pure drive process $\hat{V}_E(t,0)$ is dominant in the total process $\hat{V}_S(t,0)=\hat{V}_E(t,0)\hat{V}_{OM}(t,0)$, and $X_c$ will finally tend to $\sqrt{2}Re\{\alpha_s\}$, where $\alpha_s$ corresponds to the classical steady state value for $\hat{a}$. In the limit $g=0$, the system will reach a static steady quantum state, with the cavity being in a coherent state, after a sufficiently long time. More significant effect of  $\hat{V}_{OM}(t,0)$ with increasing coupling introduces the periodic oscillation patterns to the function $X_c(t)$, after this observable becomes stable. Given different detuning of the external driver, the stably oscillating quantum states $\rho_s(t)$ of OMS differ greatly in strong coupling regime. This can be shown by the quadrature functions $X_c(t)=Tr_S\{\rho_s(t)(\hat{a}+\hat{a}^{\dagger})/\sqrt{2}\}$ in Fig. 3.

\begin{figure}[h!]
\vspace{-0cm}
\centering
\epsfig{file=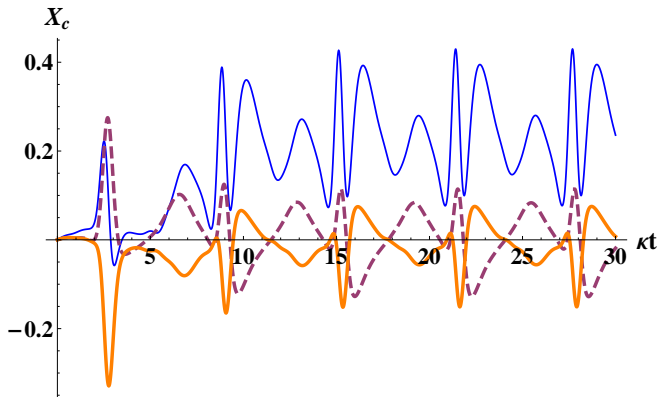,width=1.0\linewidth,clip=} 
{\vspace{-0cm}\caption{\label{Fig:m} (color online) Real-time evolution of cavity quadrature $X_c$ in strong coupling regime. The thin solid (blue) line shows the oscillation of cavity quadrature for $\Delta_0/\omega_m=0$;  the thick solid (orange) line for the detuning $\Delta_0/\omega_m=1$; the dashed line for the detuning $\Delta_0/\omega_m=-1$.
Here the system parameters are $E/\kappa=0.01$, $g/\kappa=2$, $\omega_m/\kappa=1$, $\omega_m/\gamma_m=100$, and $T=0$.}}
\vspace{-0.3cm}
\end{figure}

\begin{figure}[h!]
\vspace{-0cm}
\centering
\epsfig{file=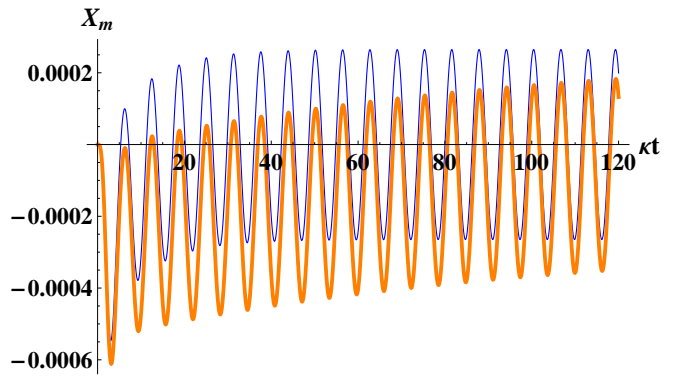,width=1\linewidth,clip=} 
{\vspace{-0.3cm}\caption{\label{Fig:m} (color online) Real-time evolution of mechanical quadrature $X_m$. The thin solid (blue) line depicts an oscillator with the quality factor $\omega_m/\gamma_m=10$, and the thick solid (orange) line for 
$\omega_m/\gamma_m=100$. The oscillation of the latter becomes stable for a longer time than the former does. The parameters are $E/\kappa=0.01$, $g/\kappa=2$, $\omega_m/\kappa=1$, and $\Delta_0/\omega_m=1$.}}
\end{figure}

\begin{figure*}[t!]
\includegraphics[width=1.0\textwidth, clip]{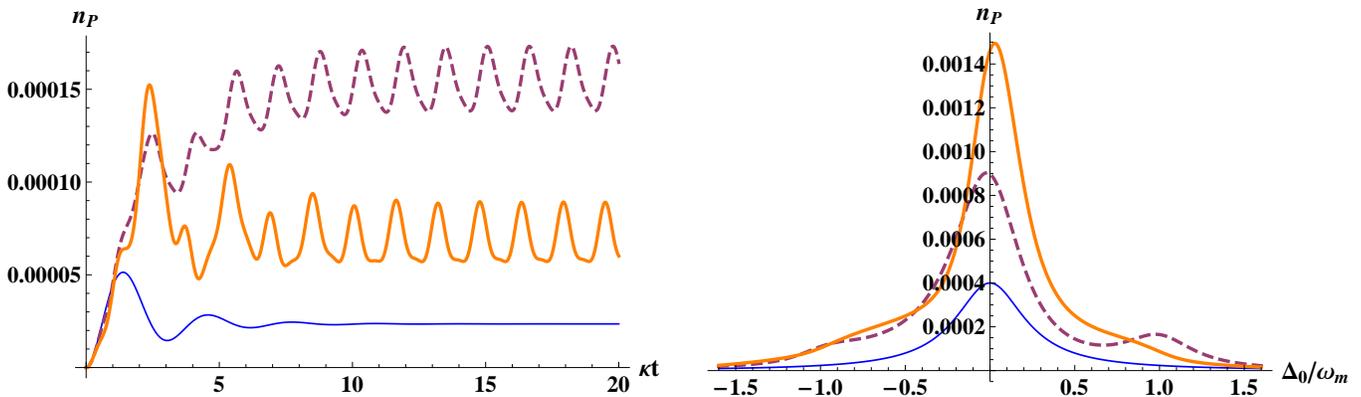}
\caption{ (color online) Left: evolution of mean cavity photon number $n_P=\langle\hat{a}^{\dagger}\hat{a}\rangle$ in strong coupling regime; Right: mean cavity photon number distribution with driver detuning at a moment. The thick solid (orange) lines represent the photon number values including the effect of cavity noise correction in (\ref{drive}); the dashed lines represent the photon number values under the approximation of the correlation function $\Gamma_c(\tau,\tau')=0$, where $\tau\neq \tau'$, in (\ref{drive}). The cavity photon number solely generated by a pure drive is shown by the thin solid (blue) lines for comparison. Here the system parameters are $E/\kappa=0.01$, $g/\kappa=2$, $\omega_m/\kappa=2$, $\omega_m/\gamma_m=100$, and $T=0$. The left frame shows the photon number evolution at the detuning point $\Delta_0=\omega_m$, and the right frame depicts the photon numbers at $\kappa t=40$. \label{cavity-photon}}
\vspace{-0cm}
\end{figure*}

The specific superposition of the harmonic components $n\omega_m$ (the first harmonic appears in Eq. (\ref{1-st})) makes the shape of a periodic pattern in Figs. 2 and 3, as its overall repetition period is controlled by the 
mechanical frequency $\omega_m$. This feature is unique to quantum OMS in strong coupling regime, and reflects the fact that stable quantum states of OMS are not static, i.e. $\dot\rho(t)\neq 0$, in the regime. Here we provide an interpretation for this result with the OMS state evolution neglecting damping to environment ($\kappa$ and $\gamma_m$ are very small to be neglected). For a zero temperature OMS driven at the detuning $\Delta_0=0$, its pure quantum state $\mbox{T}e^{-i\int_0^t\hat{H}_S(\tau) d\tau}|0\rangle_{c,m}$ under such condition can be approximated by the state 
\begin{eqnarray}
&&\mbox{T}\exp\big\{E(e^{-i\hat{\phi}_m(t)}\int_0^t  e^{i\hat{\phi}_m(\tau)}e^{i\chi_m(t,\tau)}d\tau)\hat{a}^{\dagger}\nonumber\\
&-& E(e^{i\hat{\phi}_m(t)}\int_0^t  e^{-i\hat{\phi}_m(\tau)}e^{-i\chi_m(t,\tau)}d\tau)\hat{a}\big\}|0\rangle_{c,m}
\label{app}
\end{eqnarray}
simply from a time varying drive,
where $\hat{\phi}_m(\tau)=\sqrt{2}g/\omega_m\{\sin (\omega_m \tau)\hat{x}_m-\cos(\omega_m\tau) \hat{p}_m\}$ and $\chi_m(t,\tau)=-(g/\omega_m)^2\sin\{\omega_m(t-\tau)\}$. In deriving the state in (\ref{app}) we have applied the decomposition in Eq. (\ref{1}) to the system Hamiltonian $\hat{H}_S(\tau)$ in (\ref{om}). The oscillating 
functions $\hat{\phi}_m$ and $\chi_m$ in the phases of Eq. (\ref{app}) give rise to all harmonic components of the quantum state. Eq. (\ref{app}) thus explains the cause for the dynamic quantum states of strongly coupled OMS.

The damping due to the considerable rates $\kappa$ and $\gamma_m$ will certainly decohere the quantum states of OMS, and it also determines the transient behaviors of OMS approaching stable phases. 
In Figs. 2 and 3 the transient behavior of $X_c$ lasts for a period in the order of $1/\kappa$. As a comparison, we give an example of the mechanical quadrature $X_m=\langle \hat{x}_m\rangle$ evolution in Fig. 4.
This quantity, which is proportional to the square of $E/\kappa$, is calculated with the averaged transformation $\hat{U}^{\dagger}(t,0)\hat{x}_m\hat{U}(t,0)$
over the initial cavity vacuum state and mechanical thermal state, as well as the associate reservoir states; see Appendix C. One sees that it takes time in the order of $1/\gamma_m$ for mechanical quadrature $X_m$ to reach stable oscillation.

\begin{figure}[h!]
\vspace{-0cm}
\centering
\epsfig{file=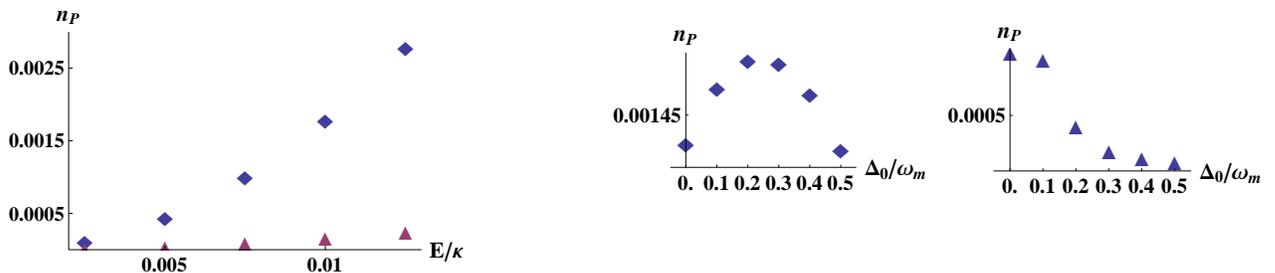,width=0.7\linewidth,clip=} 
{\vspace{-0cm}\caption{\label{Fig:m} (color online) Comparison of the cavity photon number values $n_P(E/\kappa)$ in strong coupling regime with the corresponding predictions in the linearization approach. The diamond points represent the averaged values in the stable phase (see the photon number evolution pattern in Fig. 5) as calculated in our approach, while the triangle points for the steady state values calculated with Eq. (6) of Ref. \cite{s2}. The parameters of the system are chosen as $g/\kappa=2$, $\omega_m/\kappa=2$, $\omega_m/\gamma_m=100$, $\Delta_0=0$, and $T=0$. }}
\vspace{-0cm}
\end{figure}

\begin{figure}[h!]
\vspace{-0cm}
\centering
\epsfig{file=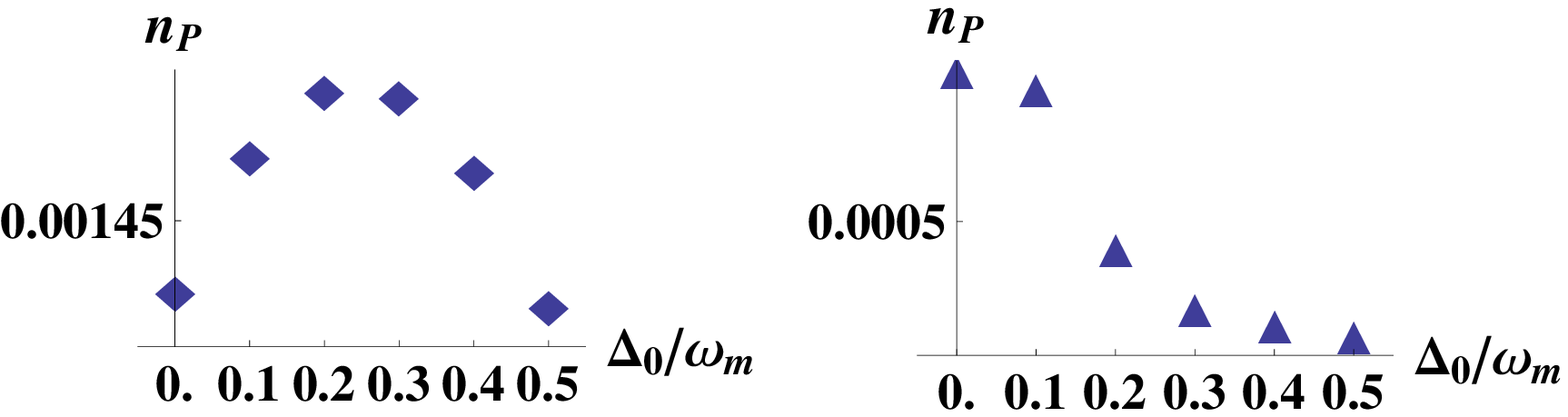,width=1.0\linewidth,clip=} 
{\vspace{-0cm}\caption{\label{Fig:m} (color online) Comparison between cavity photon numbers in bad and good cavity regime given different driver detuning. The left are the photon number values for $\omega_m/\kappa=0.5$, and the right 
for $\omega_m/\kappa=3$. Here the system parameters are $E/\kappa=0.01$, $g/\kappa=2$, $\omega_m/\gamma_m=100$, and $T=0$. These photon numbers are obtained at $\kappa t=10$.}}
\vspace{-0cm}
\end{figure}

\subsection{Cavity photon number in strong coupling regime}

The dynamics of cavity field can be clarified further by the decomposition of the system-reservoir evolution $\hat{U}(t,0)=\hat{V}_E(t,0)\hat{V}_{OM}(t,0)\hat{V}_D(t,0)$. This decomposition helps to determine the contributions to a system observable by different physical factors.
Below we study the effects of the decomposed processes on the mean cavity photon number $n_P=\langle \hat{a}^{\dagger}\hat{a}\rangle$. The system-reservoir coupling process $\hat{V}_D(t,0)$ modifies $n_P$ as it affects the system operators $\hat{a}$ and $\hat{b}$ in $\hat{V}^{\dagger}_S(t,0)\hat{a}\hat{V}_{S}(t,0)$ when acting on it; such effect exists for OMS at any temperature $T>0$ though it is not a major factor. The photon number $n_P$ is primarily determined by the pure drive process $\hat{V}_E(t,0)$ and the optomechanical coupling process $\hat{V}_{OM}(t,0)$. These main factors are described in Fig. 5 showing the evolution of the photon 
number $n_P$ at $T=0$ and its snapshot for different driver detuning at a particular moment. Similar to the evolution of cavity quadrature, mean cavity photon number tends to a steady periodic oscillation after a period in the order of $1/\kappa$; see the left frame of Fig. 5. The interference term from the two processes  $\hat{V}_E(t,0)$ and $\hat{V}_{OM}(t,0)$ contributes to the asymmetry of $n_P$ in the blue and red detuning regime, as shown in the right frame of Fig. 5. 
The coupling process $\hat{V}_{OM}(t,0)$ increases $n_P$ from that of a pure drive process $\hat{V}_E(t,0)$. For the example depicted in the figure, it magnifies the cavity photon number by more than three times around the zero-detuning resonance. 
This is different from the displaced oscillator picture in \cite{s1,s2}, where the optomechanical coupling suppresses the photon number $n_P$ at the point instead. The significant deviation between the cavity photon numbers calculated in our approach and in the linearization approach \cite{s1,s2} is shown more completely in Fig. 6, which plots the photon numbers as the function of the driving intensity $E/\kappa$. Our approach predicts a much more significant optomechanical coupling effect corresponding to the second term in Eq. ({\ref{cavity}).

Contrary to the mechanical noise effect under the condition $\gamma_m\ll\kappa$, cavity noise considerably affects system observables in strong coupling regime. Such noise corrects the pure drive operator $\hat{G}(t,\tau)$ in the process of optomechanical coupling; see (\ref{drive}). The correction takes effect as long as there exists a time window in which the correlation function $\Gamma_c(\tau,\tau')$ of the induced cavity colored noise $\hat{n}_c$ does not vanish. In Fig. 5 we compare the numerically calculated $n_P$ with both $\Gamma_c(\tau,\tau')= 0$ and $\Gamma_c(\tau,\tau')\neq 0$ for the drive operator in (\ref{drive}). The mean cavity photon numbers in the two situations differ along the most of detuning range.
Prominently the cavity photon number resonance around the red detuning point $\Delta_0=\omega_m$ and under the approximation $\Gamma_c(\tau,\tau')= 0$ is smoothed out by the cavity noise.   

Another interesting phenomenon in strong coupling regime is the shift of cavity photon number resonance from the zero detuning to the red detuned side. Given the same ratio $g/\kappa$, such shift will appear in the bad cavity regime with $\omega_m/k<1$; see the example in Fig. 7. This could be explained with the mechanical oscillator displacement in proportion to $g/\omega_m=(g/\kappa)\times (\omega_m/\kappa)^{-1}$, which is larger in bad cavity so that the effective cavity resonance frequency will be shifted to a smaller value (note that the cavity resonance frequency is inversely proportional to cavity size).

\section{Conclusion}
We have studied the dynamics of OMS weakly driven in strong coupling regime. The damping of cavity field and mechanical oscillator is treated with stochastic Hamiltonian for the coupling between system and reservoir \cite{b1}, which is also recently applied to study OMS in weak coupling regime \cite{milburn}. Different from most previous studies, the properties of OMS described here are about the situations when they are genuine quantum objects, for example, a quantum mechanical oscillator strongly coupled to a cavity single photon, rather than their classical steady states plus quantum fluctuations. For the systems in such regimes, we illustrate the dynamical evolution of cavity quadrature and mean cavity photon number under weak continuous-wave drive, as well as the significant cavity noise effect accompanying strong optomechanical coupling. The method we provide is straightforwardly applicable to OMS driven by pulses acting for limited evolution time. We expect that the dynamics of general quantum systems with strong nonlinearity and in dissipative environment could be better understood in our approach as well.

\begin{acknowledgements}
The author thanks Q. Lin for technical support and R. Ghobadi, D. Hu, F. Marquardt, L. Tian for helpful discussions. 
\end{acknowledgements}

\begin{widetext}
\section*{APPENDIX} 

\vspace{0cm}  

\subsection{Decomposition of Time-ordered Exponential}
 \renewcommand{\theequation}{A-\arabic{equation}}
 \setcounter{equation}{0}
 
In this appendix we provide an intuitive proof for Eq. (\ref{1}) and a formal proof based on differential equation for Eq. (\ref{2}). Eq. (\ref{2}) can be proved by the first method as well.

The unitary operator $\mbox{T}\exp \{-i\int_0^t d\tau \big(\hat{H}_1(\tau)  +\hat{H}_2 (\tau)\big)\}$ is the infinite product of the small elements $\hat{U}(t_i)=\exp\{-i\hat{H}_1(t_i)\delta t-i\hat{H}_2 (t_i)\delta t \}$,
where $\delta t=\lim_{N\rightarrow \infty}t/N$ and $0\leq t_i \leq t$. Within the small period $\delta t$, the small element $\hat{U}(t_i)$ can be decomposed into $\hat{U}_2(t_i)\hat{U}_1(t_i)=\hat{U}_1(t_i)\hat{U}_2(t_i)$, where $\hat{U}_i(t_k)=\exp\{-i\hat{H}_i(t_k)\delta t\}$, for any pair of $\hat{H}_1(t)$ and $\hat{H}_2(t)$, because the cross terms due to their non-commutativity are negligible. Then we will have the following expression:
\begin{eqnarray}
&&\mbox{T}\exp\big \{-i\int_0^t d\tau \big (\hat{H}_1(\tau) +\hat{H}_2 (\tau)\big)\big\}\nonumber\\
&= & \hat{U}_2(t_{N-1})\hat{U}_1(t_{N-1})\hat{U}_2(t_{N-2})\hat{U}_1(t_{N-2})\cdots\hat{U}_2(t_{2})\hat{U}_1(t_{2})\hat{U}_2(t_{1})\hat{U}_1(t_{1})\hat{U}_2(t_{0})\hat{U}_1(t_{0})\nonumber\\
&=& \hat{U}_2(t_{N-1})\hat{U}_1(t_{N-1})\hat{U}^{\dagger}_2(t_{N-1})\nonumber\\
&\times & \underbrace{\hat{U}_2(t_{N-1})\hat{U}_2(t_{N-2})}\limits_{\hat{V}_2(t,t_{N-2})}\hat{U}_1(t_{N-2})\underbrace{\hat{U}_2^{\dagger}(t_{N-2})\hat{U}^{\dagger}_2(t_{N-1})}\limits_{\hat{V}^{\dagger}_2(t,t_{N-2})}\nonumber\\
&\times & \underbrace{\hat{U}_2(t_{N-1})\hat{U}_2(t_{N-2})\hat{U}_2(t_{N-3})}\limits_{\hat{V}_2(t,t_{N-3})}\hat{U}_1(t_{N-3})\underbrace{ \hat{U}^{\dagger}_2(t_{N-3})\hat{U}_2^{\dagger}(t_{N-2})\hat{U}_2^{\dagger}(t_{N-1})}\limits_{\hat{V}^{\dagger}_2(t,t_{N-3})}\nonumber\\
&\times &\cdots \nonumber\\
&\times & \underbrace{\underbrace{\hat{U}_2(t_{N-1})\hat{U}_2(t_{N-2})\cdots \hat{U}_2(t_{1})\hat{U}_2(t_{0})}\limits_{\hat{V}_2(t,0)}\hat{U}_1(t_0)
\underbrace{\hat{U}_2^{\dagger}(t_{0})\hat{U}_2^{\dagger}(t_{1})\cdots \hat{U}^{\dagger}_2(t_{N-2})\hat{U}^{\dagger}_2(t_{N-1})}\limits_{\hat{V}^{\dagger}_2(t,0)}}\limits_{\exp\{-i\hat{V}_2(t,0)\hat{H}_1(t_0)\hat{V}^{\dagger}_2(t,0)\delta t\}}\nonumber\\
&\times & \underbrace{\hat{U}_2(t_{N-1})\hat{U}_2(t_{N-2})\cdots \hat{U}_2(t_{1})\hat{U}_2(t_{0})}\limits_{\hat{V}_2(t,0)}.
\label{infinite} 
\end{eqnarray}
Except for the bottom row, each row after the last equals sign in the above expression is the small element $\exp\{-i\hat{V}_2(t,t_k)\hat{H}_1(t_k)\hat{V}^{\dagger}_2(t,t_k)\delta t\}$. Combining these small elements as an infinite product gives the decomposition formula in (\ref{1}).

The time-ordered exponentials $\mbox{T}\exp \{-i\int_0^t d\tau \big(\hat{H}_1(\tau)  +\hat{H}_2 (\tau)\big)\}$ and $\mbox{T}\exp \{-i\int_0^t d\tau \hat{H}_1(\tau) \}$ are the solutions to the differential equations $d\hat{U}/dt=-i\big(\hat{H}_1(t)+\hat{H}_2 (t)\big)\hat{U}(t)$ and $d\hat{V}_1/dt=-i\hat{H}_1(t)\hat{V}_1(t)$, respectively. The initial condition for the differential equations is $\hat{U}(0)=\hat{V}_1(0)=I$, the identity operator.
We define $\hat{W}(t)=\hat{V}_1^{\dagger}(t)\hat{U}(t)$. Its differential with respect to $t$ reads
\begin{eqnarray} 
\frac{d\hat{W}}{dt}&=&-\hat{V}_1^{\dagger}\frac{d\hat{V}_1}{dt}\hat{V}_1^{\dagger}\hat{U}+\hat{V}_1^{\dagger}\frac{d\hat{U}}{dt}=i\hat{V}_1^{\dagger}\hat{H}_1\hat{V}_1\hat{V}_1^{\dagger}\hat{U}-i\hat{V}_1^{\dagger}(\hat{H}_1+\hat{H}_2)\hat{U}=-i\hat{V}_1^{\dagger}\hat{H}_2\hat{V}_1\hat{V}_1^{\dagger}\hat{U}=-i\hat{V}_1^{\dagger}\hat{H}_2\hat{V}_1\hat{W}.
\end{eqnarray}
The solution of the above differential equation is $\hat{W}(t)=\mbox{T}\exp \{-i\int_0^t d\tau \hat{V}_1^{\dagger}(\tau)\hat{H}_2(\tau){V}_1(\tau)\}$, implying the decomposition formula in (\ref{2}).
Note that the abbreviations $\hat{V}_i(t,0)\equiv \hat{V}_i(t)$ and $\hat{W}(t,0)\equiv \hat{W}(t)$ are used here.

\subsection{System Operator Transformation under System-reservoir Coupling} 
 \renewcommand{\theequation}{B-\arabic{equation}}
 \setcounter{equation}{0}
 
We now prove the system operator transformation under $\hat{V}_D$, the process of system-reservoir coupling in the decomposition of the overall system-reservoir evolution $\hat{U}(t,0)=\hat{V}_E(t,0)\hat{V}_{OM}(t,0)\hat{V}_D(t,0)$. The general form of this operator transformations is $\hat{V}_D^{\dagger}(\tau_2,\tau_1)\hat{c}_i \hat{V}_D(\tau_2,\tau_1)$, with $\hat{V}_D(\tau_2,\tau_1)=\hat{V}_D(t_{N-1})\cdots \hat{V}_D(t_1)\hat{V}_D(t_0)$,
an infinite product of the infinitesimal elements $\hat{V}_D(t_k)$ in the range $\tau_2\geq t_k\geq \tau_1$.
Because we apply the decomposition of Eq. (\ref{1}) in the procedure, we will find the transformation $\hat{V}_D(\tau_2,\tau_1)\hat{c}_i \hat{V}^{\dagger}_D(\tau_2,\tau_1)$ as in (\ref{1st}), and the derivation for $\hat{V}_D^{\dagger}(\tau_2,\tau_1)\hat{c}_i \hat{V}_D(\tau_2,\tau_1)$ is similar.
Let us start with the transformation by an infinitesimal element. 
To do the transformation, one expands $\hat{V}_D(t_k)$ out to second order in the stochastic increment \cite{b1}:
\begin{eqnarray}
\hat{V}_D(t_k)\hat{c}_i\hat{V}^{\dagger}_D(t_k)
&=&\exp\{\sqrt{\gamma_i} d\hat{B}_i^{\dagger}(t_k)\hat{c}_i-\sqrt{\gamma_i}d\hat{B}_i(t_k)\hat{c}_i^{\dagger}\}~\hat{c}_i~\exp\{-(\sqrt{\gamma_i}d\hat{B}_i^{\dagger}(t_k)\hat{c}_i-\sqrt{\gamma_i}d\hat{B}_i(t_k)\hat{c}_i^{\dagger})\}\nonumber\\
&=& \hat{c}_i-[\sqrt{\gamma_i}d\hat{B}_i(t_k)\hat{c}_i^{\dagger} ,\hat{c}_i]+\frac{1}{2}[\sqrt{\gamma_i}d\hat{B}_i^{\dagger}(t_k)\hat{c}_i-\sqrt{\gamma_i}d \hat{B}_i(t_k)\hat{c}_i^{\dagger},[\sqrt{\gamma_i}d\hat{B}_i^{\dagger}(t_k)\hat{c}_i-\sqrt{\gamma_i}d \hat{B}_i(t_k)\hat{c}_i^{\dagger},\hat{c}_i]]\nonumber\\
&=& (1-\frac{1}{2}\gamma_i dt)\hat{c}_i+\sqrt{\gamma_i}d \hat{B}_i(t_k),
\end{eqnarray}
where the Ito's table
\begin{eqnarray}
&& d\hat{B}_i^{\dagger}(t_k)d\hat{B}_i(t_k)=n_{th}dt,~~~d\hat{B}_i(t_k)d\hat{B}^{\dagger}_i(t_k)=(n_{th}+1)dt \nonumber\\
&& d\hat{B}_i(t_k)d\hat{B}_i(t_k)=0,~~~~~~~~~d\hat{B}_i^{\dagger}(t_k)d\hat{B}_i^{\dagger}(t_k)=0
\end{eqnarray}
has been used. 
From this small increment we will obtain the transformation 
$\hat{V}_D(\tau_2,\tau_1)\hat{c}_i \hat{V}^{\dagger}_D(\tau_2,\tau_1)$ by the following accumulation pattern:
\begin{eqnarray}
&& \hat{V}_D(\tau_2,\tau_1)\hat{c}_i \hat{V}^{\dagger}_D(\tau_2,\tau_1)=\hat{V}_D(t_{N-1})\cdots \hat{V}_D(t_{1})\hat{V}_D(t_{0}) \hat{c}_i \hat{V}^{\dagger}_D(t_{0})\hat{V}^{\dagger}_D(t_{1})\cdots\hat{V}^{\dagger}_D(t_{N-1})\nonumber\\
&=&\hat{V}_D(t_{N-1})\cdots \hat{V}_D(t_{1})\big((1-\frac{1}{2}\gamma_i\delta t)\hat{c}_i+\sqrt{\gamma_i} d \hat{B}_i(t_0)\big)\hat{V}^{\dagger}_D(t_{1})\cdots\hat{V}^{\dagger}_D(t_{N-1})\nonumber\\
&=& \hat{V}_D(t_{N-1})\cdots \hat{V}_D(t_{2})\{(1-\frac{1}{2}\gamma_i \delta t)^2\hat{c}_i+(1-\frac{1}{2}
\gamma_i \delta t)\sqrt{\gamma_i} d\hat{B}_i(t_1)+\sqrt{\gamma_i} d \hat{B}_i(t_0)\}\hat{V}^{\dagger}_D(t_{2}) \cdots \hat{V}^{\dagger}_D(t_{N-1})\nonumber\\
&=& (1-\frac{1}{2}\gamma_i \delta t)^N \hat{c}_i+(1-\frac{1}{2}\gamma_i \delta t)^{N-1}\sqrt{\gamma_i} d \hat{B}_i(t_{N-1})+\cdots+(1-\frac{1}{2}\gamma_i \delta t)\sqrt{\gamma_i}d \hat{B}_i(t_{1}) +\sqrt{\gamma_i}d \hat{B}_i(t_0)\nonumber\\
&=& e^{-\gamma_i(\tau_2-\tau_1)/2}\hat{c}_i+\sqrt{\gamma_i}\int_{\tau_1}^{\tau_2} e^{-\gamma_i(t'-\tau_1)/2}\hat{\xi}_i(t')dt',
\label{s-r}
\end{eqnarray}
where $dB_i(t)=d\int_0^t \xi_i(\tau) d\tau=\xi_i(t)dt$, and $\lim_{N\rightarrow \infty}(1-\frac{1}{2}\gamma_i \delta t)^N=e^{-\gamma_i(\tau_2-\tau_1)/2}$. Given the white noise operator $\hat{\xi}_i$ satisfying
$[\hat{\xi}_i(\tau), \hat{\xi}^{\dagger}_i(\tau')]=\delta(\tau-\tau')$, the commutation relation $[\hat{c}_i,\hat{c}_i^{\dagger}]=1$ is preserved under the transformation of $\hat{V}_D(\tau_2,\tau_1)$.
The commutator of the term $\sqrt{\gamma_i}\int_{\tau_1}^{\tau_2} e^{-\gamma_i(t'-\tau_1)/2}\hat{\xi}_i(t')dt'$
with its conjugate does not take the form of delta function, so it can be regarded as a colored noise operator.

\subsection{System Operator Transformations under Optomechanical Coupling}  
\renewcommand{\theequation}{C-\arabic{equation}}
 \setcounter{equation}{0}

The system operator transformations under $\hat{V}_{OM}(t,0)$ in the decomposition of the overall system-reservoir evolution $\hat{U}(t,0)=\hat{V}_E(t,0)\hat{V}_{OM}(t,0)\hat{V}_D(t,0)$ will be derived in this appendix. Here we use
the notation $\hat{V}_{OM}(t,0)=\mbox{T}\exp\{-i\int_0^t d\tau \hat{H}_{OM}(t,\tau)\}$, with  
 \begin{eqnarray}
\hat{H}_{OM}(t,\tau)&=&-g\big(\cos(\omega_m \tau)\hat{X}_m(t,\tau)
+ \sin(\omega_m \tau)\hat{P}_m(t,\tau) \big)\hat{C}^{\dagger}(t,\tau)\hat{C}(t,\tau)\nonumber\\
&= & -g\hat{K}_m(t,\tau) \big (e^{-\kappa (t-\tau)}\hat{a}^{\dagger}\hat{a}+e^{-\kappa (t-\tau)/2}\hat{a}\hat{G}^{\dagger}(t,\tau)+e^{-\kappa (t-\tau)/2}\hat{a}^{\dagger}\hat{G}(t,\tau)+\hat{G}^{\dagger}(t,\tau)\hat{G}(t,\tau)\big).
\end{eqnarray}

The two infinitesimal transformations for determining the transformation $\hat{V}^{\dagger}_{OM}(t,0)\hat{a}\hat{V}_{OM}(t,0)$ are
\begin{eqnarray}
e^{i\hat{H}_{OM}(t,\tau_i)\delta \tau}~\hat{a}~e^{-i\hat{H}_{OM}(t,\tau_i)\delta \tau}=\big (1+ig\hat{K}_m(t,\tau_i)e^{-\kappa (t-\tau_i)}\delta \tau\big) \hat{a}+ig\hat{K}_m(t,\tau_i)e^{-\kappa (t-\tau_i)/2}\hat{G}(t,\tau_i)\delta \tau
\label{aa}
\end{eqnarray}
and
\begin{eqnarray}
&&e^{i\hat{H}_{OM}(t,\tau_i)\delta \tau}\hat{G}(t,\tau_j)e^{-i\hat{H}_{OM}(t,\tau_i)\delta \tau}\nonumber\\
&=& \big (1+ig\Gamma_c(\tau_i,\tau_j)\hat{K}_m(t,\tau_i)\hat{G}(t,\tau_i)\hat{G}^{-1}(t,\tau_j)\delta \tau\big)\hat{G}(t,\tau_j)+ig\Gamma_c(\tau_i,\tau_j)e^{-\kappa(t-\tau_i)/2}\hat{K}_m(t,\tau_i)\hat{a}\delta \tau,
\label{GG}
\end{eqnarray}
where $\Gamma_c(\tau_i,\tau_j)=e^{-\kappa|\tau_i-\tau_j|/2}-e^{-\kappa(t-\tau_i)/2}e^{-\kappa(t-\tau_j)/2}$.
Also there is the commutation relation
\begin{eqnarray}
[\hat{K}_m(t,\tau_i),\hat{K}_m(t,\tau_j)]&=&[\cos(\omega_m \tau_i)\hat{X}_m(t,\tau_i)
+ \sin(\omega_m \tau_i)\hat{P}_m(t,\tau_i),\cos(\omega_m \tau_j)\hat{X}_m(t,\tau_j)
+ \sin(\omega_m \tau_j)\hat{P}_m(t,\tau_j)]\nonumber\\
&=& 2ie^{-\gamma_m|\tau_i-\tau_j|/2} \sin \omega_m(\tau_j-\tau_i)\equiv im(\tau_i,\tau_j)
\label{KK}
\end{eqnarray}
due to the non-commutativity of $\hat{X}_m(t,\tau)$ and $\hat{P}_m(t,\tau)$.

One of the advantages in our approach of combined system-reservoir unitary evolution is the availability of the property  $\hat{U}^{\dagger}(t,0)f(\hat{c}_i,\hat{c}_i^{\dagger})\hat{U}(t,0)=f\big(\hat{U}^{\dagger}(t,0)\hat{c}_i\hat{U}(t,0),\hat{U}^{\dagger}(t,0)\hat{c}_i^{\dagger}\hat{U}(t,0)\big)$ of a unitary transformation. This property enables one to obtain a formal expression for the transformation $\hat{V}^{\dagger}_{OM}(t,0)\hat{a}\hat{V}_{OM}(t,0)$ as follows:
\begin{eqnarray}
&&e^{i\hat{H}_{OM}(t,\tau_{N-1})\delta \tau}\cdots e^{i\hat{H}_{OM}(t,\tau_{1})\delta \tau}e^{i\hat{H}_{OM}(t,\tau_{0})\delta \tau}~\hat{a}~ e^{-i\hat{H}_{OM}(t,\tau_{0})\delta \tau}e^{-i\hat{H}_{OM}(t,\tau_{1})\delta \tau} \cdots e^{-i\hat{H}_{OM}(t,\tau_{N-1})\delta \tau}\nonumber\\
&=&  \big (1+ig\hat{K}^C_m(t,\tau_0)e^{-\kappa (t-\tau_0)}\delta \tau\big)\big (1+ig\hat{K}^C_m(t,\tau_1)e^{-\kappa (t-\tau_1)}\delta \tau\big) \cdots\big (1+ig\hat{K}^C_m(t,\tau_{N-1})e^{-\kappa (t-\tau_{N-1})}\delta \tau\big)\hat{a}\nonumber\\
&+& ig\big (1+ig\hat{K}^C_m(t,\tau_0)e^{-\kappa (t-\tau_0)}\delta \tau\big ) \cdots \big (1+ig\hat{K}^C_m(t,\tau_{N-2})e^{-\kappa (t-\tau_{N-2})}\delta \tau\big)\hat{K}^C_m(t,\tau_{N-1})\hat{G}_C(t,\tau_{N-1})\delta \tau\nonumber\\
&+& \cdots +ig\big (1+ig\hat{K}^C_m(t,\tau_0)e^{-\kappa (t-\tau_0)}\delta \tau\big) e^{-\kappa (t-\tau_{1})/2}\hat{K}^C_m(t,\tau_{1})\hat{G}_C(t,\tau_{1})\delta \tau+ig e^{-\kappa (t-\tau_{0})/2}\hat{K}^C_m(t,\tau_{0})\hat{G}_C(t,\tau_{0})\delta \tau\nonumber\\
&=& \big(\mbox{T}\exp\{-ig\int_0^t d\tau e^{-\kappa (t-\tau)}\hat{K}^C_m(t,\tau)\}\big)^{\dagger}\hat{a}\nonumber\\
&+&ig\int_0^t d\tau~ 
\big (\mbox{T}\exp\{-ig\int_0^{\tau} du~ e^{-\kappa (t-u)}\hat{K}^C_m(t,u)\}\big)^{\dagger} e^{-\kappa(t-\tau)/2}\hat{K}^C_m(t,\tau)
\hat{G}_C(t,\tau),
\label{formal}
\end{eqnarray}
where $\hat{K}^C_m(t,\tau_i)
= \hat{V}_{OM}^{\dagger}(t,\tau_i)\hat{K}_m(t,\tau_i)\hat{V}_{OM}(t,\tau_i)$, and
$\hat{G}_C(t,\tau_i)
= \hat{V}_{OM}^{\dagger}(t,\tau_i)\hat{G}(t,\tau_i)\hat{V}_{OM}(t,\tau_i)$.

To apply (\ref{formal}) to numerical calculations, one should find the proper forms of $\hat{K}^C_m(t,\tau)$ and $\hat{G}_C(t,\tau)$.
The modified operator $\hat{K}^C_m(t,\tau_i)$ is obtained by the successive infinitesimal unitary operations (from $\tau_{i+1}$ to $t$) on $\hat{K}_m(t,\tau_i)$, which comes from the infinitesimal transformation $e^{i\hat{H}_{OM}(t,\tau_i)\delta \tau}~\hat{a}~e^{-i\hat{H}_{OM}(t,\tau_i)\delta \tau}$ at the moment $\tau_i$; see the following:
\begin{eqnarray}
\hat{K}^C_m(t,\tau_i)&=&e^{i\hat{H}_{OM}(t,\tau_{N-1})\delta \tau }\cdots e^{i\hat{H}_{OM}(t,\tau_{i+1})\delta \tau}\hat{K}_m(t,\tau_i) e^{-i\hat{H}_{OM}(t,\tau_{i+1})\delta \tau }\cdots e^{-i\hat{H}_{OM}(t,\tau_{N-1}\delta \tau }\nonumber\\
&=& \hat{K}_m(t,\tau_i)+g\int_{\tau_i}^t d\tau~ m(\tau,\tau_i)\hat{V}_{OM}^{\dagger}(t,\tau)\hat{C}^{\dagger}(t,\tau)\hat{C}(t,\tau)\hat{V}_{OM}(t,\tau).
\label{m-k}
\end{eqnarray}
The above equation is expanded to
\begin{eqnarray}
\hat{K}^C_m(t,\tau_i)&=&\hat{K}_m(t,\tau_i)+g\int_{\tau_i}^t d\tau~ e^{-\kappa (t-\tau)}m(\tau,\tau_i)
\hat{V}_{OM}^{\dagger}(t,\tau)\hat{a}^{\dagger}\hat{a}\hat{V}_{OM}(t,\tau)\nonumber\\
&+& g\int_{\tau_i}^t d\tau~ e^{-\kappa (t-\tau)/2}m(\tau,\tau_i)\hat{V}_{OM}^{\dagger}(t,\tau)\big(\hat{a}\hat{G}^{\dagger}(t,\tau)+\hat{a}^{\dagger}\hat{G}(t,\tau)\big)\hat{V}_{OM}(t,\tau)\nonumber\\
&+&g\int_{\tau_i}^t d\tau~ m(\tau,\tau_i)\hat{V}_{OM}^{\dagger}(t,\tau)\hat{G}^{\dagger}(t,\tau)\hat{G}(t,\tau)\hat{V}_{OM}(t,\tau).
\label{m-correction}
\end{eqnarray}
The term carrying the decayed correlation function $e^{-\kappa (t-\tau)}m(\tau,\tau_i)$ in the integrand can be neglected. 

Next, based on the infinitesimal transformation (\ref{GG}), one will also find 
\begin{eqnarray}
\hat{G}_C(t,\tau_i)
&=& \hat{G}(t,\tau_i)+ig\int_{\tau_i}^t d\tau~\Gamma_c(\tau,\tau_i)\hat{V}_{OM}^{\dagger}(t,\tau)\hat{K}_m(t,\tau)\hat{V}_{OM}(t,\tau)\hat{V}_{OM}^{\dagger}(t,\tau)\hat{G}(t,\tau)\hat{V}_{OM}(t,\tau)\nonumber\\
&+&ig\int_{\tau_i}^t d\tau~e^{-\kappa(t-\tau)/2}\Gamma_c(\tau,\tau_i)\hat{V}_{OM}^{\dagger}(t,\tau)\hat{K}_m(t,\tau)\hat{V}_{OM}(t,\tau)\hat{V}_{OM}^{\dagger}(t,\tau)\hat{a}\hat{V}_{OM}(t,\tau).
\label{c-ca}
\end{eqnarray}
The last term with the decayed correlation function $e^{-\kappa(t-\tau)/2}\Gamma_c(\tau,\tau_i)$ in the integrand can be well neglected, and the remaining terms will be expanded by the iteration of the above to
\begin{eqnarray}
\hat{G}_C(t,\tau_i)&=&\hat{G}(t,\tau_i)+ig\int_{\tau_i}^t d\tau \Gamma_c(\tau,\tau_i)\hat{K}^C_m(t,\tau)\hat{G}(t,\tau)\nonumber\\
&+&(ig)^2 \int_{\tau_i}^t d\tau \Gamma_c(\tau,\tau_i)\hat{K}^C_m(t,\tau)\int_{\tau}^t d\tau' \Gamma_c(\tau',\tau)\hat{K}^C_m(t,\tau')\hat{G}(t,\tau')\nonumber\\
&+& (ig)^3 \int_{\tau_i}^t d\tau \Gamma_c(\tau,\tau_i)\hat{K}^C_m(t,\tau)\int_{\tau}^t d\tau' \Gamma_c(\tau',\tau)
\hat{K}^C_m(t,\tau')
\int_{\tau'}^t d\tau'' \Gamma_c(\tau'',\tau')\hat{K}^C_m(t,\tau'')\hat{G}(t,\tau'')+\cdots.\nonumber\\
\label{G-C}
\end{eqnarray}

In principle, the negligence of the integrals carrying the decayed correlation functions for Eqs. (\ref{m-correction}) and (\ref{c-ca})
is the only approximation made in our procedure. For a point $\tau_i$ outside the vicinity of the end time point $t$, the decay factor $e^{-\kappa (t-\tau)/2}$ completely damps the correlation function $\Gamma_c(\tau,\tau_i)$ inside the correlation time window around it to zero, so there is no contribution from the neglected integral. 
The correlation function $m(\tau,\tau_i)$ defined in (\ref{KK}) takes the form of oscillating function for large quality factor $\omega_m/\gamma_m\gg 1$, suppressing the integral of the positive term $\hat{V}_{OM}^{\dagger}(t,\tau)\hat{a}^{\dagger}\hat{a}\hat{V}_{OM}(t,\tau)$ even without the damping factor $e^{-\kappa (t-\tau)}$.

More approximations could be made to simplify the above expressions. 
First, the summation of multi-folded integrals in (\ref{G-C}) can be conveniently estimated given a short correlation time window of the colored cavity noise, in which the drive operator $\hat{G}(t,\tau)$ changes slowly. In this way such summation can be approximated by a time-ordered exponential. Second, for a weak drive of $E/\kappa\ll 1$, the operator $\hat{K}^C_m(t,\tau_i)$ can be simply approximated by $\hat{K}_m(t,\tau_i)$. Putting these approximations together, one will have the following closed form of the transformation 
\begin{eqnarray}
&&\hat{V}^{\dagger}_{OM}(t,0)\hat{a}\hat{V}_{OM}(t,0)\nonumber\\
&=& e^{ig\int_0^t d\tau e^{-\kappa (t-\tau)}\sin(\omega_m\tau)\hat{P}_m(t,\tau)}e^{ig\int_0^t d\tau e^{-\kappa (t-\tau)}\cos(\omega_m\tau)\hat{X}_m(t,\tau)}e^{-ig^2\Theta(t)}\hat{a}\nonumber\\
&+& ig \int_0^t d\tau  e^{-ig^2\Theta(\tau)}e^{ig\int_0^\tau du e^{-\kappa (t-u)}\sin(\omega_m u)\hat{P}_m(t,u)}e^{ig\int_0^\tau du e^{-\kappa (t-u)}\cos(\omega_m u)\hat{X}_m(t,u)}\nonumber\\
&\times & e^{-\kappa(t-\tau)/2}\hat{K}_m(t,\tau)\big \{e^{ig\int_\tau^t du \Gamma_c(u,\tau)\sin(\omega_m u)\hat{P}_m(t,u)}e^{ig\int_\tau^t du \Gamma_c(u,\tau)\cos(\omega_m u)\hat{X}_m(t, u)}e^{-ig^2\Theta'(\tau)}\hat{G}(t,\tau)\big\}
\label{closed-form}
\end{eqnarray}
for a single-photon weak driver. Here we also have the extra phases 
$$\Theta(\tau)=2\int_0^\tau du e^{-\kappa (t-u)}\sin(\omega_m u)\int_0^{u} dv e^{-\kappa(t-v)-\gamma_m (u-v)/2}\cos(\omega_m v), $$
$$\Theta'(\tau)=2\int_{\tau}^{t} du~ \Gamma_c(u,\tau)\sin(\omega_m u)\int_{\tau}^u dv ~\Gamma_c(v,\tau)e^{-\gamma_m(u-v)/2}\cos(\omega_m v),$$ 
after factorizing the time-ordered exponentials involving $\hat{X}_m(t,u)$ and $\hat{P}_m(t,u)$.

Moreover, the pure driving operation on cavity operator is simply found as
\begin{eqnarray}
\hat{V}^{\dagger}_{E}(t,0)\hat{a}\hat{V}_{E}(t,0)&=&\mbox{T}\exp\big \{E\int_0^t d\tau  (\hat{a}e^{-i\Delta_0 \tau}e^{-\frac{\kappa}{2}(t-\tau)}-h.c.)\big\}~\hat{a}~ \mbox{T}\exp\big \{E\int_0^t d\tau  (\hat{a}^{\dagger}e^{i\Delta_0 \tau}e^{-\frac{\kappa}{2}(t-\tau)}-h.c.)\big\}\nonumber\\
&=& \hat{a}+E\int_0^t d\tau~e^{i\Delta_0 \tau}e^{-\frac{\kappa}{2}(t-\tau)}=\hat{a}+D_1(t).
\end{eqnarray}

Finally, the transformations of the mechanical oscillator operators $\hat{U}^{\dagger}(t,0)\hat{x}_m\hat{U}(t,0)$ and $\hat{U}^{\dagger}(t,0)\hat{p}_m\hat{U}(t,0)$, where 
$\hat{U}(t,0)=\hat{V}_E(t,0)\hat{V}_{OM}(t,0)\hat{V}_D(t,0)$, can be found in a similar way to (\ref{m-k}). 
Under the same approximation as for (\ref{closed-form}), their averages over the initial state of cavity vacuum, thermal state of mechanical oscillator, as well as the cavity vacuum reservoir and mechanical thermal reservoir state, read
\begin{eqnarray}
\langle\hat{U}^{\dagger}(t,0)\hat{x}_m\hat{U}(t,0)\rangle &=& -\sqrt{2}g\int_0^t d\tau~ e^{-\gamma_m (t-\tau)/2}\sin(\omega_m\tau)\big(e^{-\kappa(t-\tau)/2}D_1^{\ast}(\tau)+D_2^{\ast}(\tau)\big)\big(e^{-\kappa(t-\tau)/2}D_1(\tau)+D_2(\tau)\big),\nonumber\\
\langle\hat{U}^{\dagger}(t,0)\hat{p}_m\hat{U}(t,0)\rangle &=&~\sqrt{2}g\int_0^t d\tau~ e^{-\gamma_m (t-\tau)/2}\cos(\omega_m\tau)\big(e^{-\kappa(t-\tau)/2}D_1^{\ast}(\tau)+D_2^{\ast}(\tau)\big)\big(e^{-\kappa(t-\tau)/2}D_1(\tau)+D_2(\tau)\big).~~~~~~~~~~
\end{eqnarray}
These averages are used to study mechanical oscillator dynamics under general optomechanical coupling.
\vspace{1cm}

\end{widetext}


\begin{thebibliography}{99}
\bibitem {ep} At the temperature $T=0$ one has more simplified averages in (\ref{1-sim}), with $\hat{V}_D(t,0)\rho(0)\otimes \rho_R \hat{V}^{\dagger}_D(t,0)=\rho(0)\otimes \rho_R$ for the initial vacuum state $\rho(0)\otimes \rho_R=|0\rangle_{c,m}\langle 0|\otimes |0\rangle_R\langle 0|$.
\end{thebibliography}
\end{document}